\documentclass[preprint]{aastex}
\usepackage{epsfig}
\usepackage{color}

\newcommand{\Htwo}{\mbox{H$_{2}$}}
\newcommand{\Jp}{$J^{\prime}$}
\newcommand{\Jpp}{$J^{\prime\prime}$}
\newcommand{\Vp}{$v^{\prime}$}
\newcommand{\Vpp}{$v^{\prime\prime}$}
\newcommand{\jp}{J^{\prime}}
\newcommand{\jpp}{J^{\prime\prime}}
\newcommand{\vp}{v^{\prime}}
\newcommand{\vpp}{v^{\prime\prime}}

\newcommand{\xsig}{$X\,^{1}\Sigma^{+}_{g}$}
\newcommand{\bsig}{$B\,^{1}\Sigma^{+}_{u}$}
\newcommand{\cpi}{$C\,^{1}\Pi_{u}$}
\newcommand{\kms}{km~s$^{-1}$}

\slugcomment{\Htwo\ $\tau(\lambda)$ Templates}

\shorttitle{Molecular Hydrogen Templates}
\shortauthors{McCandliss }

\begin{document}

%\include{draft.tex}

\title{Molecular Hydrogen Optical Depth Templates for {\em FUSE} Data Analysis}


\author{S. R. McCandliss,\altaffilmark{1} }
\affil{Department of Physics and Astronomy, The Johns Hopkins University,
Baltimore, Maryland 21218}
\email{stephan@pha.jhu.edu}


\altaffiltext{1}{Department of Physics and Astronomy, The Johns Hopkins University,
Baltimore, Maryland 21218}


\begin{abstract}

The calculation and use of molecular hydrogen optical depth templates
to quickly identify and model molecular hydrogen absorption features
longward of the Lyman edge at 912 \AA\ are described.  Such features
are commonly encountered in spectra obtained by the Far Ultraviolet
Spectroscopic Explorer and also in spectra obtained by the Space
Telescope Imaging Spectrograph, albeit less commonly.  Individual
templates are calculated containing all the Lyman and Werner
transitions originating from a single rotational state (\Jpp) of the
0th vibrational level (\Vpp) of the ground electronic state. Templates
are provided with 0.01~\AA\ sampling for doppler parameters ranging
from 2~$\le b \le$~20~km~s$^{-1}$ and rotational states 0~$\le \jpp
\le$~15.  Optical depth templates for excited vibrational states are
also available for select doppler parameters.  Each template is calculated for a
fiducial column density of log[N(cm$^{-2}$)]~=~21 and may be scaled to
any column less than this value without loss of accuracy.  These templates will facilitate the determination of
the distribution of molecular hydrogen column density as a function of
rotational level. The use of these templates will free the user
from the computationally
intensive task of calculating  profiles for a large number of
lines and allow 
concentration on line profile
or curve-of-growth fitting  to determine column densities and doppler
parameters.
The
templates may be downloaded freely from the URL
http://www.pha.jhu.edu/$\sim$stephan/h2ools2.html.

\end{abstract}




\section{Introduction}

The launch of the Far Ultraviolet Spectroscopic Explorer {\em FUSE} has
opened a new spectral window on the universe with a 905~--~1187~\AA\
bandpass \citep{Sahnow2000}.  Within this window spectra of continuum emitting objects
almost universally exhibit absorption bands of molecular hydrogen,
indicating this gas resides in the interstellar medium somewhere
between the object and the observer.  These bands arise from the
photo-excitation of molecular hydrogen by the background continuum
object, causing electrons in the ground electronic state
$X\,^{1}\Sigma^{+}_{g}$, to be excited into either the higher
$B\,^{1}\Sigma^{+}_{u}$ (Lyman band), or  $C\,^{1}\Pi_{u}$ (Werner
band) states.  
The strength and density of the absorption lines are proportional to 
the distribution of molecular hydrogen column densities
among the ground state rotational and vibrational
(\Jpp, \Vpp) energy levels\footnote[2]{When referring to the rotational and
vibrational quantum numbers in a transition going from an upper
electronic state (e.g. either \bsig\ or \cpi) to the ground state
(\xsig) it is traditional to designate the upper state numbers with a
single prime (e.g. \Jp, \Vp) and the lower state with a double prime
(e.g.  \Jpp, \Vpp).} peculiar to the
line-of-sight.

For some observers the molecular hydrogen electron population
distribution within the ro-vibrational states is a subject of
direct interest.  For others, molecular hydrogen absorption is a nuisance,
potentially contaminating or even obscuring spectral features of interest.  Fortunately there
is a high degree of correlation between the absorption lines within
different ro-vibrational bands (\Vp--~0) imposed by the uniformity of
the energy level separations and the slow monotonic variation of
oscillator strength as a function of upper vibrational state, a
consequence of the Franck-Condon principle \citep{Herzberg1950}.  This
correlation greatly aids the identification of molecular hydrogen
features and provides a means for their removal from blended features,
provided they can be reliably modeled in an unobscured portion of the
spectrum.

Here we will discuss the calculation and use of a simple set
of optical depth templates from which observers can easily identify and model
the molecular hydrogen features commonly observed in the {\em FUSE}
spectral range.  Each template contains all the Lyman and Werner
absorption lines 
arising from the photo-excitation of electrons out of     
common ground state
rotational  \Jpp\ and vibrational  \Vpp\ = 0 levels (usually) and into 
the \Vp\ states of the upper  bands
(\Vp~$\leftarrow$~0), subject to the appropriate rotational state selection rules.  A
single column density $N{(\vpp\!,\jpp)}$ controls the line strength and the
templates are scaleable with this quantity.  The optical depth
variation with wavelength for each line is calculated using a Voigt
profile to allow for the proper variation of line shape with increasing
column density, reproducing the shapes endemic to the linear, flat, and
square root portions of the curve-of-growth as controlled through the
specification of the doppler parameter ($b$).  The templates are given
as optical depths functions, as opposed to transmission functions, to
facilitate the inclusion of various atomic and molecular species or
multiple velocity structures, simply by adding additional optical
depths to the template function at the appropriate wavelength
position.

These templates can be used in conjunction with a variety of methods to
determine the appropriate column densities and doppler parameters.
One  method is direct profile fitting, where the templates for various
$N{(\vpp\!,\jpp)}$ and $b$ are used with a predetermined continuum to
directly compare to the observed spectrum in a $\chi^2$ calculation.
Another method uses the curve-of-growth technique, where equivalent
width measurements are used to determine the appropriate column and
doppler parameter from which the absorption line transmission function
can be quickly calculated.  Once the columns, doppler parameters and
continua are known in an unblended portion of the spectrum the degree to which
they effect the other features, such as \ion{O}{6} or \ion{D}{1}, maybe assessed.

The template calculations incorporate  the
Abgrall et al.  (1993a,b) databases of transition
lifetimes, wave-numbers, rotational and vibrational quantum numbers for
Lyman and Werner band transitions.  The templates are
available for free access on-line through
\verb+http://www.pha.jhu.edu/~stephan/h2ools2.html+. Here we provide
potential users with a ready reference to the calculation procedures,
some examples of use, and discuss appropriate positional accuracy and
column density limitations.  We include formulae for
the calculation of energy levels, population distributions, and line
profiles in Appendices A and B.

\section{Energy Levels,  Population Distribution, and Selection Rules}

The ground state rotational energy levels in the \Vpp~=~0 vibrational
state have temperature equivalent energies $B_{v}hc/k~\approx$~0, 170,
510, 1014, 1680, 2500 K for  $J~=~0~\rightarrow~5$, while  the
vibrational energy levels are much more widely spaced with separations
$\Delta~G^{\prime\prime}hc/k~\equiv~(G(\vpp+1)-G(\vpp))(hc/k)~\lesssim$~5983~K
(see Appendix A).  Boltzmann's law shows that the first few rotational
states are low enough to be populated at the relatively cold
temperatures associated with molecular gas in the interstellar medium,
while excitation of the upper vibrational levels requires a more
energetic environment.  Absorption from several rotational levels is
the norm, although the early {\it Copernicus} observations showed the
typical distribution of column density in the higher rotational states
is larger than expected from a single temperature Boltzmann
distribution \citep{Aannestad1973}.  The reasons for the deviations
from a Boltzmann distribution are myriad and we refer to the reader to
the comprehensive review by \citet{Shull1982} who have discussed the
various diagnostics for determining formation rates, collision rates
and  local radiation field intensity by using the rotational column
density distribution  of molecular hydrogen and the atomic hydrogen
column.   Absorption from excited vibrational states are rare.  When
present they indicate that the gas is affected by either a high
temperature process such as shocks \citep{McCandliss2001} or a
nonthermal processes such as fluorescent pumping \citep{Meyer2001}.

Lyman bands have two branches R(\Jpp), with \Jp~=~\Jpp~+~1 and P(\Jpp),
with \Jp~=~\Jpp~-~1.  The selection rule that requires $\Delta
J$~=~$\pm$~1 results from the zero change in electronic angular
momentum  between the \xsig\ and \bsig\ states \citep{Herzberg1950}.
The Werner bands have an additional branch Q(\Jpp) with  $\Delta
J$~=~0, allowed because the change in electronic angular momentum
between the \xsig\ and \cpi\ states is one.  In addition, the minimum
\Jp\ for the \bsig\ state is 0, while the minimum \Jp\ for the
\cpi\ state is 1.  This leads to an absence of P(0) in the Lyman bands
and P(0), P(1) and Q(0) in the Werner bands.

There are 377 absorption line transitions from Lyman
(0~$\le~\vp~\lesssim~20$) and Werner (0~$\le~\vp~\lesssim~5$) bands
arising from rotational levels, 0~$\le \jpp \le 6$ in the ground
vibrational level (\Vpp=~0) of the \xsig\ electronic state longward of
the Lyman edge.  Many more are possible if \Vpp~$>$~0 becomes
populated.  Consequently the identification of the molecular hydrogen
lines can be daunting. It is typically the first task in the post
pipeline analysis of {\em FUSE} spectra.  The identification process is
eased considerably by locating the strongest Lyman lines  usually R(0)
for \Jpp~=~0  and  P(1),  R(1) for \Jpp~=~1, and then proceeding to
higher \Jpp\ until the lines merge with the noise.  Our computation of
templates containing only those lines originating from a single
rotational state allows for a direct implementation of this method.  It
greatly facilitates the determination of the $N(\vpp,\jpp)$
distribution, and allows for assessing the degree of contamination in
coincident alignments of molecular hydrogen absorptions with lines from
other diagnostically useful atoms or molecules.

\section{Optical Depth Template Calculation}

A total optical depth function is constructed for the \Jpp\ rotational
state of the \Vpp\~=~0 vibrational level from the sum of all the
individual optical depth profiles  ranging over all \Vp\ states in the
Lyman (0~$\le~\vp~\le$~18) and Werner (0~$\le~\vp~\le$~5) line database
given as, 
\begin{equation}
\tau_{\jpp}(\lambda)~=~~\Sigma_{\vp=0}^{v_{max}} \tau_i(\lambda).
\end{equation} 
The function $\tau_{\jpp}(\lambda)$ is established on a
common wavelength grid spanning the region of interest
(790~--~1490~\AA\ in our case) with 0.01~\AA\ spacing.  This total
optical depth function is used to accumulate individual optical depth
profiles.  Each line profile is calculated using the Voigt function
formulae in Appendix B, covering a range of wavelengths spanning
50~\AA\ to either side of the line center,  so as to accommodate the
broad Lorentzian wings that appear in the line profiles at high column
density.  The span of the common wavelength grid is arbitrary.  It
allows for the convenient inclusion of contributions from lines of
absorbing species both inside and outside of the {\em FUSE} bandpass.
An insertion point is determined to be the wavelength grid point
closest to the line center and the individual optical depth profile is
added to the total optical depth about that point and the process is
repeated for all Lyman and Werner transitions arising from the same
lower level rotational state.

For the \Jpp~=~0 states the template contains only the R(0) lines of
the  Lyman  and Werner bands of (\Vp~$\leftarrow$~0).  For \Jpp~$=$~1
there are R(1) and P(1) lines in the Lyman  bands and R(1) and Q(1)
lines in the Werner bands.  Above \Jpp~$=$~1 there are R(\Jpp) and
P(\Jpp) lines in the Lyman  bands and R(\Jpp), Q(\Jpp) and P(\Jpp)
lines in the Werner bands.  A representative set of templates for
0~$\le~\jpp~\le$~4 with a doppler parameter of 5~\kms\ with log columns of
21 and 20  for \Jpp~=~0,1, 19 to 18 for \Jpp~=~2,
17 for \Jpp~=~3, and 16 for \Jpp~=~4.
are displayed in Figure \ref{fig1}.

\section{Line Position, Equivalent Width and Column Density Limits}

The location of each line within the larger wavelength grid is accurate
to within the grid step or 3~\kms. This method of insertion is
computationally expedient compared to the more accurate but much slower
method of interpolating the centered profile onto the 0.01~\AA\ grid.
Our grid spacing is on the order of the largest differences between
observed and calculated wave-numbers in the Abgrall et al.  (1993a,b)
databases ($\sim$ 1 cm$^{-1}$).  The  resolution of the {\em FUSE}
spectrograph is $R~\approx$~20000 or 15~\kms\ \citep{Sahnow2000}, so the grid
oversamples by a factor of 5.

The minimum grid step of 0.01~\AA\ sets a lower limit on the useful
doppler parameter range.  The core of a line with a doppler parameter
of 1~\kms\ is not sampled with sufficient resolution to yield an
accurate measure of  optical depth with increasing distance from the
line core.  Consequently, the numerically integrated equivalent width
from the 1~\kms\ template is higher than the theoretical value in the
linear portion of the curve-of-growth and lower in the flat portion of
the curve-of-growth.  Although available on the website, the
1~\kms\ templates  should not be used for quantitative
work.   This problem is quantified in Figure~\ref{fig2} where we show
curves-of-growth (in red) for a range of $Nf\lambda$, and a descrete
set of doppler parameters (b=1, 2, 4, 8, and 16 \kms). Overplotted are
the equivalent widths derived from direct numerical integration of the
Lyman lines in the J=0 and J=1 templates over the same range of doppler
parameters.  The problem with the equivalent widths calculated from the
lines contained in the b~=~1  \kms\ template is evident.  There is a
slight  ($<$~7\%) residual from the theoretical curves in 
b~=~2~\kms\ derived equivalent widths.

An examination of the current literature shows that very few
lines-of-sight have been found with molecular hydrogen determined
doppler parameters $<$~2~\kms.  For example, in a study of the Large
and Small Magellanic clouds \citet{Tumlinson2002} find  only two
lines-of-sight (out of 37 determinations) that have  molecular hydrogen
doppler velocities of $<$~2~\kms, although several have lower bounds
that dip below b~=~2~\kms.  Another study by \citet{Meyer2001} finds a
b~=~1.8~($\pm$0.1)~\kms\ for the highly excited molecular hydrogen in
the line-of-sight toward HD 37903, the central star of the reflection
nebula NGC~2023.   These instances notwithstanding, the current set of
templates should be adequate for most work.  We note that
lines-of-sight with b~=~2~\kms\ will have instrumental line profiles,
except where column densities are high enough to produce damping wings. The curve-of-growth
technique will be prefered to line profile fitting for determining
column densities and doppler parameters in these instances, as noted in
the following section.  The templates are not used in
the curve-of-growth analysis except to represent the solution after it
has been found.  The regime where absorption lines will have instrumental
profiles in the FUSE spectrographs is roughly identified in
Figure~\ref{fig2} as the region below the solid horizontal line drawn
at $\log{W/\lambda}$~=~1/$R$.

Nevertheless there are many cases where small doppler velocities have
been measured in species that correlate well with molecular hydrogen,
as in the case of CH (e.g. \citet{Andersson2002} and
\citet{Federman1982}).  In order to support such investigations in the
future, templates will be developed with finer wavelength and doppler
velocity grids for exploring the low b regime.  Higher resolution grids
will also find use in studies requiring closely spaced multiple
velocity components with small doppler velocities.  However, the
current set of grids are large ($\approx$ 8Mbyte uncompressed for a
single b up to \Jpp~=~15) and the high resolution grids will be at
least twice as large.

The column densities have no lower limits, they may be scaled to as low
a column as required without loss of accuracy. An upper limit results
from the finite wavelength span of $\pm$50~\AA\ for the individual
profile calculations.  Above $\log{N(\vpp\!,\,\jpp)}$~=~21 (the
fiducial column density for which each optical depth template is
calculated) the individual line profiles suffer  a progressively severe
truncation of the Lorentz wings at $\pm$50~\AA\ from line center,
which  manifests itself as a series of discontinuities in the summed
optical depth grid.  The $\log{N(\vpp\!,\,\jpp)}$~$\lesssim$~21 is a
``fuzzy limit'' as the seriousness of the truncation depends upon the
oscillator strength.  This  effect is independent of doppler parameter
because at $\log{N(\vpp\!,\,\jpp)}$~$\ge$~21 all lines are fully damped
for b~$<$~20~\kms.

The effect is illustrated in  Figure~\ref{fig3} with the lines of
minimum and maximum oscillator strength in the R(0) transitions of the
Lyman bands (0 -- 0) and (7 -- 0),  where we show the half profiles of
these two lines over a 50 \AA\ span.  We see that for a column density
of $\log{N(0,0)}$~=~21 the line profile has returned from black at the
line core to 0.9995 and 0.9990 at 50 \AA\ from line center for (0 -- 0)
and (7 -- 0), respectively.  For a column density of
$\log{N(0,0)}$~=~22 the lines only return to 0.995 and 0.990, while for
$\log{N(0,0)}$~=~23 the return is 0.95 and 0.90, respectively.  It is
difficult to estimate quantitatively the combined effects of profile
truncations since the  error it generates is a function of wavelength.
The general effect is to cause a progressive underestimation of the
combined blackness of the rotational template for column densities
$\log{N(\vpp\!,\,\jpp)}$~$\ge$~21. Quantitative work at columns much
greater than $\log{N(\vpp\!,\,\jpp)}$~$\ge$~21 using these templates
should be avoided.  This should not be too much of a problem as few
sight lines with  $\log{N(\vpp\!,\,\jpp)}$~$>$~21 have been found.  For
instance, in an exploration of 23 sight lines with $A_V~\gtrsim$~1
\citet{Rachford2002} found only 1 case (HD~154368) where the formal
value of $\log{N(\jpp=0)}$ was 21.04~$\pm$~0.05.

\section{Discussion and Examples}

The templates are stored in files with names that encode the calculation parameters.  
For example, tauh2n21b2j0-15v0.dat, indicates a column density of $\log{N(cm^{-2})}$~=~21, a doppler width of 2 km s$^{-1}$, a rotational
state range of of  0~$\le~\jpp\le~$15, and a ground vibrational state of
\Vpp~=~0.  The files contain  17 arrays of unformatted double precision numbers with 59000 elements.  The first array is the wavelength variable. The next 16 arrays are the \Jpp~=0, 1, $\cdots$ 15 optical depth functions.  Table 2 gives a
schematic of how
the data arrays are placed in the file. A simple IDL program is provided
on the website to
read in the data and form an example transmission function. 
There are extra empty elements longward of the {\em FUSE} long wavelength cutoff at 1189~\AA\  in part because there are a few high rotational lines that exist beyond this cutoff.  However, the main reason  is to allow for the inclusion
of spectra for ground state vibrational levels $\vpp~>$~0, for the purpose of investigating molecular hydrogen excitation processes in energetic astrophysical
environments.  Rotational templates for $\vpp~>$~0 states will be included on the H$_2$ools website in the near future.

Absorption spectra (the transmission functions) are obtained by taking
e$^{\tau_{\lambda}}$.  All the optical depth templates are tabulated as negative numbers.  The spectra for
column densities other than 10$^{21}$ particles cm$^{-2}$ can be
obtained by dividing the optical depth by 10$^{21}$ and multiplying by
the required column density value. For example, the absorption spectrum for 10$^{18}$
is given by e$^{\tau_{\lambda}/1000}$.  These spectra should be
convolved  with an instrumental line spread function to simulate
instrumental broadening before comparing to an  object spectrum.  For
use with {\em FUSE} spectra, convolution with a gaussian width with a FWHM on
the order $R~\gtrsim$~15000 is typical. For high precision work it may
be necessary to employ a wavelength dependent convolution kernel, as
the resolution of the various spectral channels is wavelength dependent
at the 10\% level \citep{Sahnow2000}.

Once convolved, an overlay of the $J^{\prime\prime}$ = 0
transmission function ( multiplied by an appropriate continuum) on an object spectrum will allow an assessment of the need for zero-point offsets to
the wavelength registration. The $J^{\prime\prime}$ = 0 spectrum contains the
bluest lines in any given Lyman band ($v^{\prime}$ -- 0), so it should line
up with the blue side of the blended R(0) and R(1) lines.  For the highly damped lines encountered at high column density it may be necessary to include higher \Jpp\ states to
assess zero-point offsets.  Adjust the column density of the
absorption spectrum  until a good fit is obtained.  Once satisfied add
the optical depth template from the next highest rotational state, take the exponent, adjust the
columns and wavelength zero-point as necessary, and reassess the fit.
Repeat until the inclusion of a higher level \Jpp\ no longer improves the fit.  The process may be quantified  by computing the $\chi^2$
statistic over  wavelength intervals selected to coincide with unblended  
absorption features in  the object spectrum.  Repeat for different column densities and 
doppler parameters and then find the minimum $\chi^2$.

It it is often expedient to perform the initial fitting operation
by eye.  Interactive assessment can be used to determine the columns
to within a few dex and the doppler parameters within a few
km~s$^{-1}$, provided complicated velocity structures on the line of
sight are absent and the column density is high enough.  These approximate determinations can serve as the starting
point for more precise work and minimize the time devoted to $\chi^2$
minimization by providing good guesses.  In Figure~\ref{fig4} we
show how the higher vibrational transitions can be used to define the
column and the lower vibrational transitions the doppler parameter when
the absorptions are making the transition from the flat portion of the
curve-of-growth to the square root portion where the lines profiles
become fully damped.  At $\log{N(\jpp=0)}$~=~19 we see little
difference in the higher vibrational line profiles for the two doppler
parameters, while the lower vibrational lines do show a difference. At low
doppler parameter the lines are still saturated, having box like
profiles, while damping wings  have developed in the higher doppler
parameter line.

In the low column transition region ($\log{N(\jpp=0)}$~$\sim$~15)
between the linear and flat portion of the curve-of-growth, discerning
the tradeoff in column and doppler parameter is less apparent to the
eye.  In this regime the lines are unresolved (by {\em FUSE}) and will have
profiles close to the instrumental profile except for the highest doppler
parameters.  In these cases the curve-of-growth method for
determination of column is generally more useful than profile fitting.
In between these two regimes  on the flat  (slowly varying) portion of
the curve-of-growth, it is relatively easy to determine the doppler parameter (b) from the box
like profiles, but the column is ill-defined.  Here profile fitting is
the most accurate method for column determination.

In Figure~\ref{fig5} we show a curve-of-growth example. The
molecular hydrogen column densities and a doppler parameter have been derived from a curve-of-growth analysis
of all the lines in the rotational states 0~$\le~\jpp~\le$~5 between 1040~--~1098~\AA.
The bottom right panel for the (4--0) Lyman band fit shows an overplot
of the transmission function computed by applying the derived column densities
and doppler parameter. 
A curve-of-growth analysis was then applied to eight \ion{D}{1} lines starting
at Lyman-$\gamma$.  An absorption line model containing lines from \ion{H}{1},
\ion{D}{1}, \ion{O}{1}, and molecular hydrogen has been overplotted in each panel.

In Figure~\ref{fig6} we show an example of a $\chi^2$ fit to the
(4--0) Lyman band in the lower panel, which was fine-tuned by taking
into account the fit to the (0--0) Lyman band.    The rotational
templates were used with an appropriate continuum to calculate $\chi^2$
for selected portions of the line profiles.   In the upper panel of we
illustrate how well the  \Htwo\ model from the (4--0) band fit
reproduces the \Htwo\ structures in the (5--0) band.  The template
model can be used to constrain the location and degree of blending with
lines of \ion{O}{6}, \ion{C}{2}, \ion{O}{1} and \ion{Ar}{1}.  Several
unidentified lines are also evident, some of which are higher
rotational level \Htwo\ lines.

These are just two examples illustrating how the use of these templates
can expedite the identification, computation and deblending of
molecular hydrogen profiles from far-UV spectra.  The use of a common
0.01 \AA\ wavelength grid facilitates the inclusion of profiles from
other atomic and molecular species.  In addition, any number of
multiple velocity components can easily be coadded and scaled, with the
appropriate doppler dilatation or contraction  applied through the
interpolation of the optical depth template onto a shifted wavelength
grid.  Such interpolated optical depth templates can be employed to search
for molecular hydrogen in intervening intergalactic absorption systems
that appear in the spectra of quasars and galaxies at high redshift.
The templates can find use in any study that requires an accounting
for molecular hydrogen blending.  Examples include studies seeking to
determine the velocity distribution and abundance of  \ion{O}{6} in the
galaxy and halo, of \ion{D}{1} in the local and nearby ISM, or of any
other interstellar atomic and molecular species that may blend with
molecular hydrogen features.  The templates also
provide a means to determine the excitation state of molecular hydrogen
so as to directly test the predictions of various excitation mechanisms,
such as fluoresence (radiative pumping), formation pumping, or shock excitation.
These templates can easily be used with $\chi^2$ minimization codes to
assess the goodness-of-fit as functions of column density and
doppler parameter.  They provide a means for the user to concentrate
on the fitting procedure without having to take on the computationally
intensive task of calculating line profiles for a large number of
absorbers.   Such goodness-of-fit codes are under development and testing by many groups (c.f. Owens.f discussed in Lemoine~et~al.~2002  and  Rachford~et~al.~2002), and some of these codes are available on-line\footnote[3]{http://origins.colorado.edu/$\sim$tumlinso/h2/sw/sw.html}.  The author will entertain inquiries into the use of beta versions that he is developing
as time allows.

\acknowledgments

I would like to acknowledge David Neufeld for the original suggestion
to calculate templates for the individual rotational states and William
Blair for organizing a series of tutorials on molecular hydrogen that
provided the impetus for this work.   I would also like to acknowledge
Brian Wolven for providing an electronic version of the Abgrall et al.
(1993a,b) databases and Ravi Sankrit for commenting on initial drafts of this paper.  The calculations described here were performed in
IDL on a unix workstation.  Some users, working on non-unix machines, have reported  problems with
byte swapping when downloading the templates and I would like to thank
Pat Romano for kindly providing a set of byte swapped templates that were developed for use
on linux machines.  Finally, simulating discussion provided by various
members of the {\em FUSE} science and operations team is greatly appreciated.
This work has been supported by NASA grants NAG5-5315, NAG5-10403, and
NAG5-11456  to the Johns Hopkins University.



\appendix
\section{Term Values and Thermal Population Distribution}

Following \citet{Herzberg1950}, the ro-vibration energy levels (in wave-number units of cm$^{-1}$) of an anharmonic oscillator is a sum of electronic, vibrational and rotational
term values,
\begin{equation}
T(v,J) = T_e + G(v) + F(v,J) 
\end{equation}
where $T_e$ is a constant for the given electronic state.  $G(v)$ and
$F(v,J)$ are given as expansions in $(v+\frac{1}{2})$ and $J(J+1)$ respectively
with $v=0, 1~\cdots~v_{max},~J=J_{min}, J_{min}+1~\cdots~J_{max}$,
\begin{eqnarray}
G(v)& = & \omega_e(v+\frac{1}{2}) - \omega_{e}x_{e}(v+\frac{1}{2})^2 + \omega_{e}y_{e}(v+\frac{1}{2})^3 + \cdots, \\
F(v,J)&=&B_{v}J(J+1) - D_{v}(J(J+1))^2 + \cdots,
\end{eqnarray}
and $B_v$ and $D_v$ are also given as expansions in $(v+\frac{1}{2})$,
\begin{eqnarray}
B_{v}&=&B_e - \alpha_e(v+\frac{1}{2}) + \cdots, \\
D_{v}&=&D_e + \beta_e(v+\frac{1}{2}) + \cdots .
\end{eqnarray}
We reproduce in Table 1 the most significant term value constants for the \xsig,~ \bsig,~ and \cpi~ states collected by
\citet{Huber1979}.    They have also compiled a comprehensive list of the higher order constants of these expansions.  
 
The \xsig\ energy levels may be used in Boltzmann's law to determine
the distribution of electrons among the ro-vibrational levels of the ground
state under the assumption of thermal equilibrium.  Following
\citet{Draine1996} we write, 
\begin{equation} N(v,J)/g_J = N~exp[\frac{-hcT(v,J)}{kt}] 
\end{equation} 
where $N$ is a
renormalization factor determined by first calculating the matrix
elements N(v,J) with N = 1 and then defining \(N = \Sigma_{v=0
J=0}^{v_{max} J_{max}} N(v,J) \) for the renormalization.  $g_J$ is the
degeneracy of the states where for even $J$ (para-H$_2$) $g_J = 2J+1 $
and for odd $J$ (ortho-H$_2$) $g_J =3(2J+1)$ under the assumption that
the ortho to para ratio is 3.

\section{Absorption Profiles}

The formulae reproduced here are quite general and may be used in the calculation of any absorption
line profile, atomic or molecular.
The shape and strength of an absorption line depends on 5 independent constants,  the oscillator strength f$_{\vp \vpp \jp \jpp}$, the
wavelength $\lambda_{\vp \vpp \jp \jpp}$, the total lifetime of the upper
state $\Gamma_{\vp \vpp \jp \jpp}$,  the doppler velocity $b~=~\sqrt{(2kT/m_{H_2})^2~+~v^2_{turb}}$, and the column density of the lower state $N{(\vpp\!,\jpp)}$. The first 3 variables are set by quantum mechanics while the latter 2 are the quantities of 
interest along the line-of-sight.  At first glance it would appear that we need to calculate separate
templates for a grid of column densities and doppler parameters. However,
this is not necessary because the cross-section profile is independent of
column density so
the optical depth 
scales linearly with column density.  Consequently a
series of optical depth templates, calculated with the same doppler velocity, can
be directly scaled with column density, reducing our need for an optical
depth series where the doppler parameter alone is varied.

Following \citet{Cartwright1970}, the normalized
line profile as a function of wavelength ($\lambda$) for the
absorbing transition of an electron in level $\vp \jp \leftarrow \vpp \jpp$ 
(which we will designate with an index $i$) maybe expressed as:
\begin{eqnarray}
I_i(\lambda)~&=&~
\exp{(-\tau_i(\lambda))}, \\
\tau_i(\lambda)~&=&~N(\vpp\!,\jpp)~\sigma_i(\lambda), \\
\sigma_i(\lambda)~&=&~\sigma_i(\lambda_i) H(a,u), \\
\sigma_i(\lambda_i)~&\equiv&~\frac{\sqrt{\pi} e^2}{m_e c b} f_i \lambda_i, \\
H(a,u)~&=&~\frac{a}{\pi}~\int^{\infty}_{-\infty}\frac{e^{-y^2} dy}{a^{2} + (u - y)^2}.
\end{eqnarray}
Here $H(a,u)$ is the Voigt function, a convolution of the a Maxwellian
velocity distribution with a Lorentz profile \citep{Rybicki1979}.  The
dimensionless frequency is $u~=~\frac{\nu-\nu_{i}}{\Delta\nu_d}$  and
$a~=~\frac{\Gamma_i}{4~\pi~\Delta\nu_d}$ is the damping parameter.  The
continuous, discrete and doppler frequencies are $\nu~=c/\lambda$,
$\nu_{i}~=c/\lambda_i$, and $\nu_d~=(b/c)\nu_{i}$ ($c$ is the speed of
light, $e$ is the electron charge in esu, $m_e$ is the electron mass in grams,
and b is defined above).  The total lifetime of the upper state is a sum over all
possible vibrational transitions out of the upper \Vp\ state and into
the lower states 0~$\le~\vpp~\le$~14 such that
\(\Gamma_i~=~\Sigma_{\vpp=0}^{14} A_i\) where
$A_i~=~\frac{2\jpp+1}{2\jp+1}\frac{8~\pi^2~e^2}{m_e~c~\lambda_i}~f_i$.
The 
line core cross-section is $\sigma_i(\lambda_i)$ as the Voigt function has a value of $H(0,0)~=~1$
at line center (when $a~<<~1$ as is usually the case).  $\sigma_i(\lambda)$
and $\tau_i(\lambda)$ are the cross-section and optical depth profiles,
respectively.  

In Figure~\ref{fig2} the curves-of-growth shown in red result
from the numeric integration of $1-\exp{(-\tau_i(\lambda))}$  
over the complete wavelength interval of the profile, for a range of column densities, and descrete
set of doppler parameters. 
The (well known) formulae for the linear, flat, and square root portion of the curves are overplotted (in black) to directly show the regions of validity
for the approximations.

\begin{eqnarray}
\frac{W_{\lambda}}{\lambda}~&=&~\frac{\pi e^2 N f \lambda}{m_e c^2} \\[.25in]
			 ~&=&~2\frac{b}{c} [ \ln{ (\frac{ \sqrt{\pi} e^2  N f \lambda}{m_e c b}) }  ]^\frac{1}{2} \\[.25in]
	~&=&~ [ \frac{ e^2~\gamma~N f \lambda^2 }{ m_e~c^3 } ]^\frac{1}{2}
\end{eqnarray}

These formulae are discussed in detail in \citep{Spitzer1978}.  
It can be seen in Figure~\ref{fig2} that the formula for the flat portion of the curve-of-growth
is low by approximately 6 -- 8\% with respect to the numeric calculation.  An examination
of \citep{Munch1968} shows this descrepancy  can be accounted for by keeping
the second order term in the asymptotic expansion of $\frac{W_{\lambda}}{\lambda}$ for large $\tau_0 = (\frac{ \sqrt{\pi} e^2  N f \lambda}{m_e c b}) \gg\ 1$.  The correction factor to $\frac{W_{\lambda}}{\lambda}$ is then $(1~+~.2886/\ln{\tau_0})$ and setting
$\tau_0 \approx\ $ 50 yields a 7\% correction.

\figcaption[]{ \label{fig1} The transmission functions as derived for all
the Lyman and Werner absorption lines originating from \Jpp~=~0,1,2,3,4
longward of 900 \AA, top to bottom respectively.  The vibrational transitions
(\Vp~--~0) are labeled above the lines with the Werner band label above
the Lyman.  The log columns are
21 and 20  for \Jpp~=~0,1, 19 to 18 for \Jpp~=~2,
17 for \Jpp~=~3, and 16 for \Jpp~=~4.  These columns were chosen to be representative of the monotonically decreasing distribrutions commonly derived from far UV spectra.}

\figcaption[]{ \label{fig2} Curve-of-growth diagram illustrating the regions of
validity for each of the equivalent width approximation formulae.  The numeric solution is in red, and the approximations are in black.   Overplotted are the equivalent
widths derived from direct numerical integration of the Lyman lines in
the J=0 and J=1 templates for b~=~1, 2, 4, 8, and 16 \kms\ showing the problem
with the   b~=~1  \kms\ template.  The solid black horizontal line at
the middle of the graph indicates roughly the boundary between  resolved
and unresolved (instrumentally broaded) line profiles, assuming the
FUSE resolution is $R$~=~20000.  The dashed lines
at the bottom of the graph indicate the 3$\sigma$ equivalent width detection
limits for continuum signal-to-noise ratios of 10 and 30 per resolution element.
The placement of these lines was
determined with the formula $W~=~3(N/S)(\lambda/R)$.}

\figcaption[]{ \label{fig3} On the right is a blowup of the Lyman R(0)(7-0) line
as inserted into the common wavelength grid showing the truncation edge
at -50 \AA\ from line center.  The $\log N$ scalings of 21, 22 and 23 are overplotted.  On the left is the same for Lyman R(0)(0-0), which
has a smaller oscillator strength.  The truncation edge is not as
deep in the line on the right. }

\figcaption[]{ \label{fig4} The variation of line profile for
a fixed column of $\log N$ = 19 in the  
R(0) (3--0), (2--0), (1--0) and (0--0) Lyman lines with two different  doppler parameter (b) of 5~km~s$^{-1}$
and 10~km~s$^{-1}$ are overplotted.  The shortward lines are very nearly the same, having become nearly fully damped they differ only in the core, while the (0--0) line for b~=~5~km~s$^{-1}$ is still on the flat portion of the curve-of-growth.  This illustrates how the higher vibrational transitions can be used to define the column and the lower vibrational transitions the doppler parameter
b when the absorptions are making the transition from the flat portion of the
curve-of-growth to the square root portion where the lines profiles become
fully damped. }

\figcaption[]{ \label{fig5} Curve-of-growth fitting example.  The
\Htwo\ column densities and a doppler parameter have been derived from a curve-of-growth analysis.
The fit for the (4--0) Lyman band fit is displayed in bottom right panel. 
An absorption line model containing lines from \ion{H}{1},
\ion{D}{1}, \ion{O}{1}, and \Htwo\ has been overplotted in each panel.
The \ion{D}{1} and \ion{O}{1} models were also derived from curve-of-growth analysis while the \ion{H}{1} model was ``$\chi$-by-eye'' from the Lyman-$\gamma$ profile, which exhibits a damping wing.}

\figcaption[]{ \label{fig6} $\chi^2$ line profile fitting example.  The rotational templates were used with an appropriate continuum
to calculate $\chi^2$ for selected portions of the line profiles
of the  (4--0) Lyman band.   
The upper panel shows 
how well the  \Htwo\ model from the (4--0) band fit 
reproduces the \Htwo\ structures in the (5--0) band and
constrains the location and
degree of blending with lines of \ion{O}{6}, \ion{C}{2}, \ion{O}{1} and
\ion{Ar}{1}.}

%% Tables should be submitted one per page, so put a  before

%% Two options are available to the author for producing tables:  the
%% deluxetable environment provided by the AASTeX package or the LaTeX
%% table environment.  Use of deluxetable is preferred.
%%

%% Three table samples follow, two marked up in the deluxetable environment,
%% one marked up as a LaTeX table.

%% In this first example, note that the \tabletypesize{}
%% command has been used to reduce the font size of the table.
%% Note also that the \label command needs to be placed 

%\clearpage

\begin{deluxetable}{rrrrrrr}
\tablecolumns{8}
\tablewidth{0pc}
\tablecaption{Molecular Hydrogen Term Value Constants}
\tablehead{
\colhead{State}   & \colhead{$T_e$}    & \colhead{$\omega_e$} &
\colhead{$\omega_{e}x_{e}$}    & \colhead{$B_e$}   & \colhead{$\alpha_e$}    & \colhead{$D_e$}}
\startdata 
\cpi\ &100089.8&2443.77&69.524&31.362&1.664&2.23$\times$10$^{-2}$\\
\bsig\ &91700.0&1358.09&20.888&20.015&	1.1845&1.625$\times$10$^{-2}$\\
\xsig\ &0&4401.21&121.33&60.853&3.062&4.71$\times$10$^{-2}$\\
\enddata
\end{deluxetable}

\clearpage

\begin{deluxetable}{cccc}
\tablecolumns{4}
\tablewidth{0pc}
\tablecaption{Template File Arrays}
\tablehead{\colhead{Variable}   & \multicolumn{3}{c}{Arrays} }
\startdata
$\lambda_i$			& $\lambda_0$ = 900	& \nodata & $\lambda_{58999}$ = 1489.99 	\\				  	
$\tau(\lambda_i)_{\jpp=0}$ 	& $\tau(\lambda_0)$	& \nodata	&$\tau(\lambda_{58999})$\\
			 	
$\vdots$		 		&  & \vdots &  \\		
$\tau(\lambda_i)_{\jpp=15}$ 	& $\tau(\lambda_0)$ & \nodata&$\tau(\lambda_{58999})$ \\

\enddata
\end{deluxetable}


\begin{thebibliography}{}

\bibitem[Aannestad \& Field(1973)]{Aannestad1973} Aannestad, P.~A.,~\& 
Field, G.~B.\ 1973, \apjl, 186, L29 

\bibitem[Abgrall et al.(1993a)]{Abgrall1993a} Abgrall, H., Roueff, 
E., Launay, F., Roncin, J.~Y., \& Subtil, J.~L.\ 1993a, \aaps, 101, 273 

\bibitem[Abgrall et al.(1993b)]{Abgrall1993b} Abgrall, H., Roueff, 
E., Launay, F., Roncin, J.~Y., \& Subtil, J.~L.\ 1993b, \aaps, 101, 323 

\bibitem[Andersson, Wannier, \& Crawford(2002)]{Andersson2002} 
Andersson, B.-G., Wannier, P.~G., \& Crawford, I.~A.\ 2002, \mnras, 334, 
327 



\bibitem[Cartwright \& Drapatz(1970)]{Cartwright1970} Cartwright, 
D.~C.,~\& Drapatz, S.\ 1970, \aap, 4, 443 

\bibitem[Draine \& Bertoldi(1996)]{Draine1996} Draine, B.~T.,~\& 
Bertoldi, F.\ 1996, \apj, 468, 269 

\bibitem[Federman(1982)]{Federman1982} Federman, S.~R.\ 1982, \apj, 
257, 125

\bibitem[Herzberg(1950)]{Herzberg1950} Herzberg, G.\ 1950, Spectra of Diatomic Molecules, New York: 
Van Nostrand Reinhold,  2nd ed. 


\bibitem[Huber \& Herzberg(1979)]{Huber1979}  Huber, K. P., \& Herzberg, G. 1979, Molecular Spectra and Molecular Structure IV: Constants of Diatomic Molecules New York: Van Nostrand-Reinhold, p. 249 -- 250

\bibitem[Lemoine et al.(2002)]{Lemoine2002} Lemoine, M.~et al.\ 
2002, \apjs, 140, 67 



\bibitem[McCandliss(2001)]{McCandliss2001}
{McCandliss}, S.~R. 2001, ASP Conf. Ser. 247: Spectroscopic Challenges of Photoionized Plasmas, ed. by Gary Ferland and Daniel Wolf Savin,  p.  523 -- 527


\bibitem[Meyer et al.(2001)]{Meyer2001} Meyer, D.~M., Lauroesch, 
J.~T., Sofia, U.~J., Draine, B.~T., \& Bertoldi, F.\ 2001, \apjl, 553, L59 

\bibitem[M\"{u}nch(1968)]{Munch1968} M\"{u}nch, G.\ 1968, Stars and Stellar Systems VII, Chicago: University of Chicago Press, p. 365 

\bibitem[Rachford et al.(2002)]{Rachford2002} Rachford, B.~L.~et 
al.\ 2002, \apj, 577, 221 



\bibitem[Rybicki \& Lightman(1979)]{Rybicki1979} Rybicki, G.~B.,~\& 
Lightman, A.~P.\ 1979, New York: Wiley-Interscience,  p. 291 

\bibitem[Sahnow et al.(2000)]{Sahnow2000} Sahnow, D.~J.~et al.\ 
2000, \apjl, 538, L7 


\bibitem[Shull \& Beckwith (1982)]{Shull1982} Shull, J.~M.,~\& 
Beckwith, S.\ 1982, \araa, 20, 163 

\bibitem[Spitzer(1978)]{Spitzer1978} Spitzer, L.\ 1978, Physical Processes
in the Interstellar Medium, New York: 
Wiley-Interscience,  p. 52  

\bibitem[Tumlinson et al.(2002)]{Tumlinson2002} Tumlinson, J.~et al.\ 
2002, \apj, 566, 857 




\end{thebibliography}
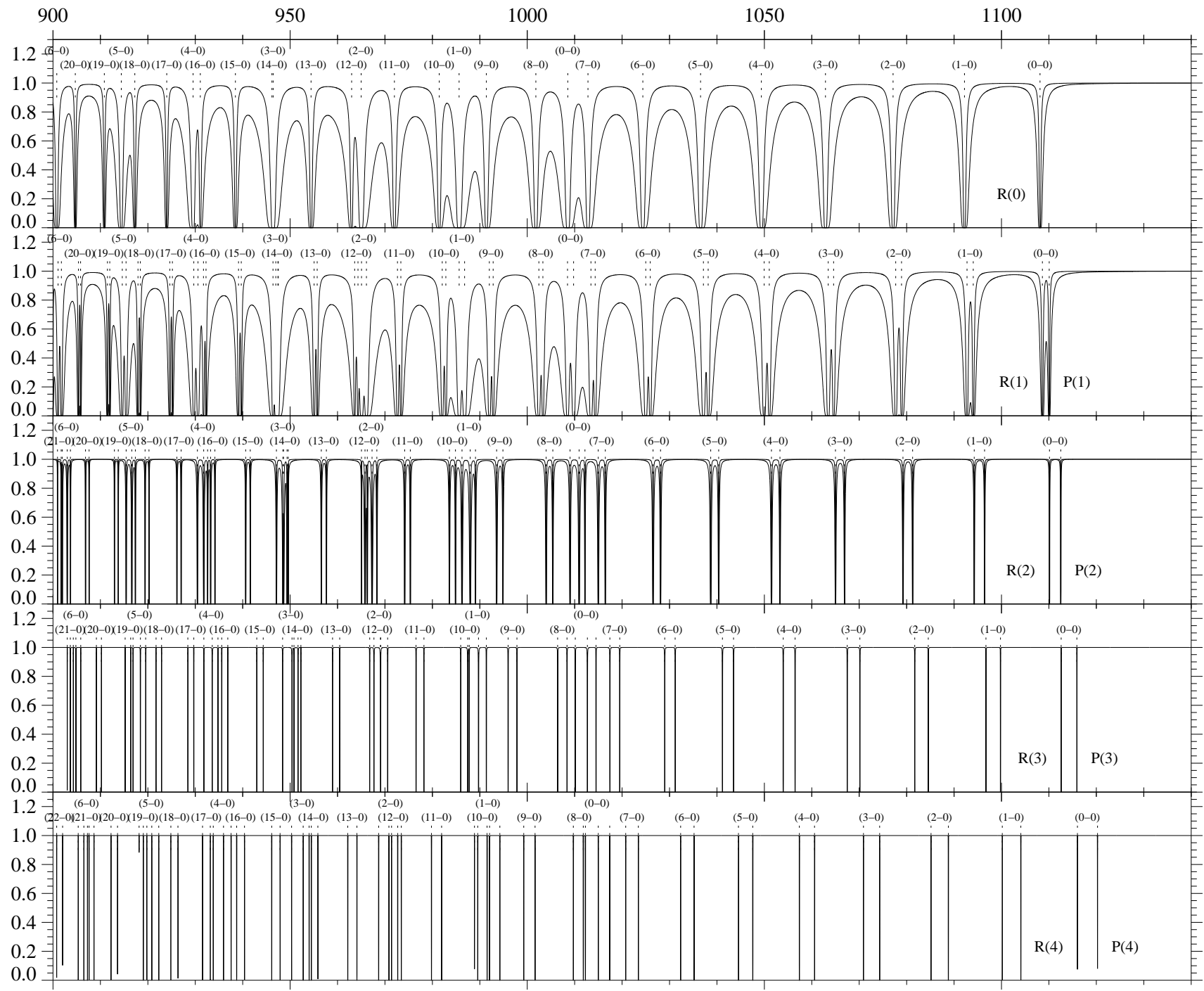
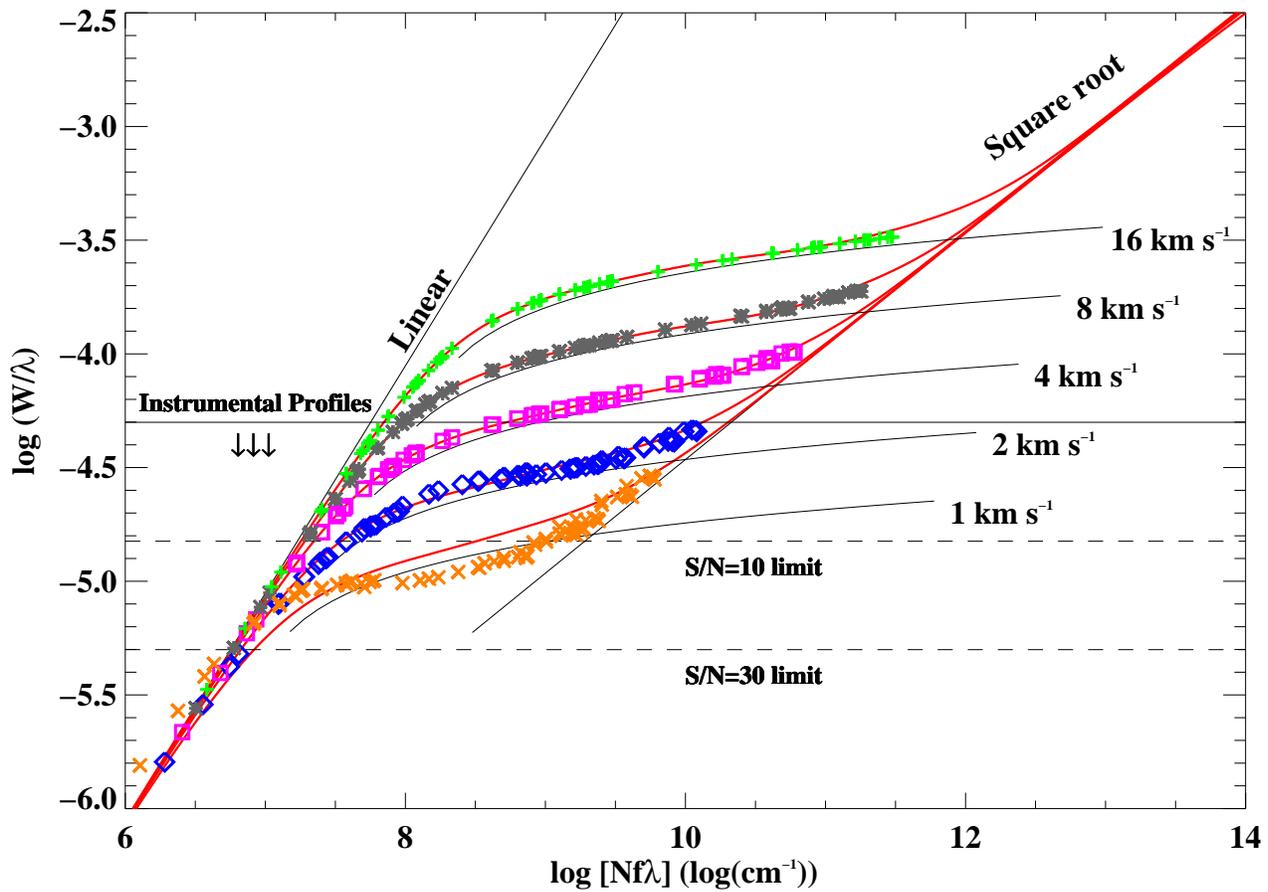
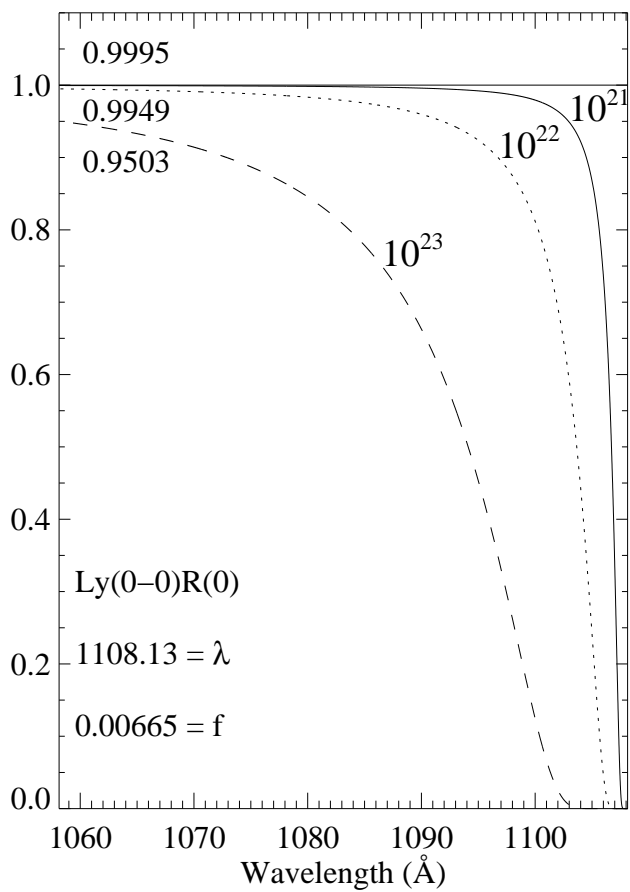
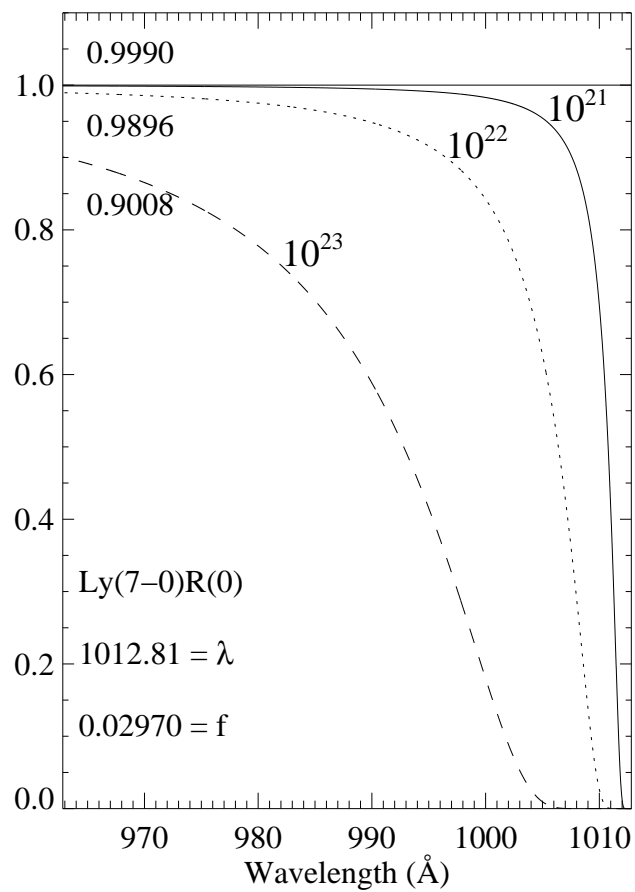
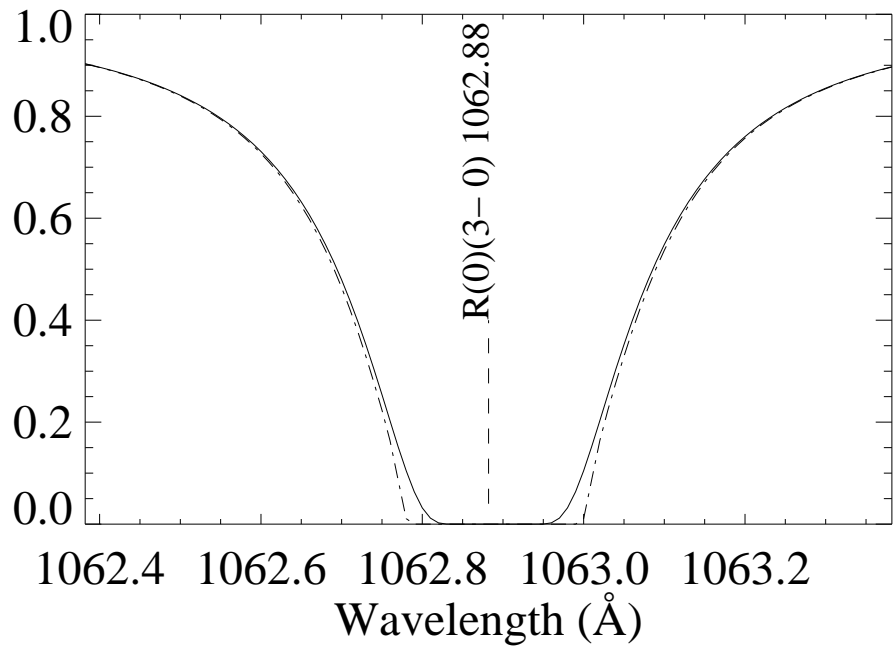
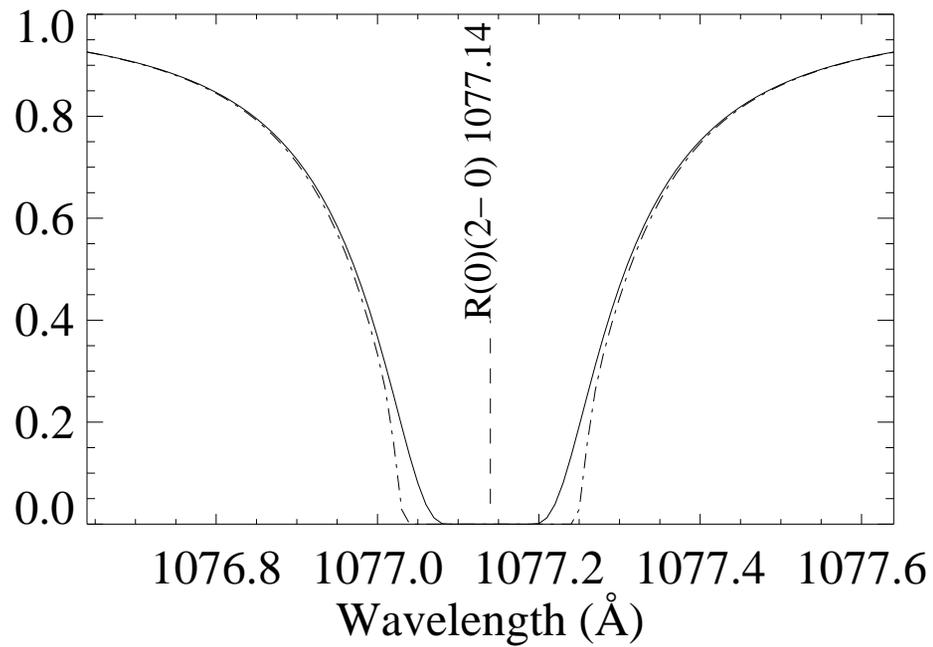
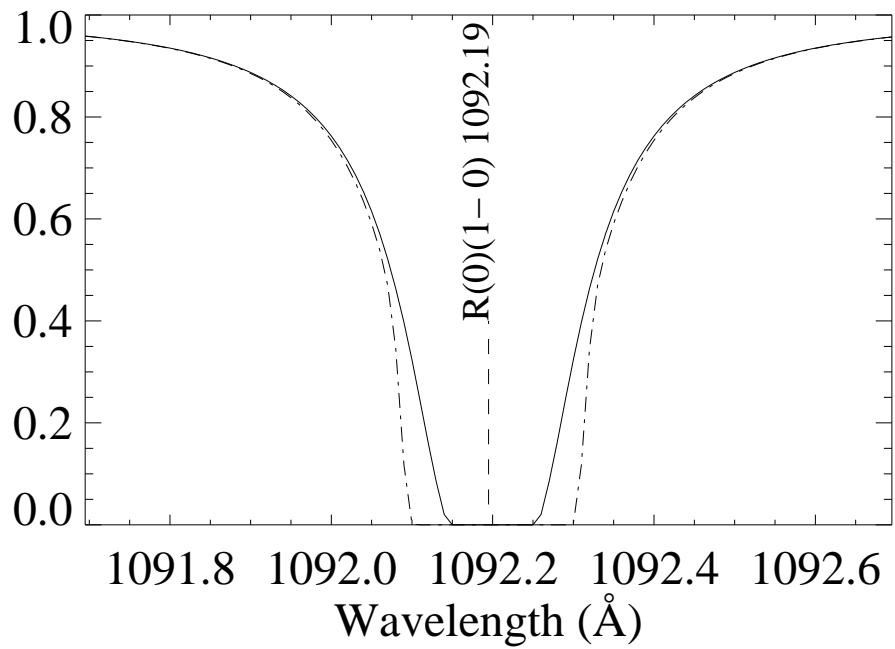
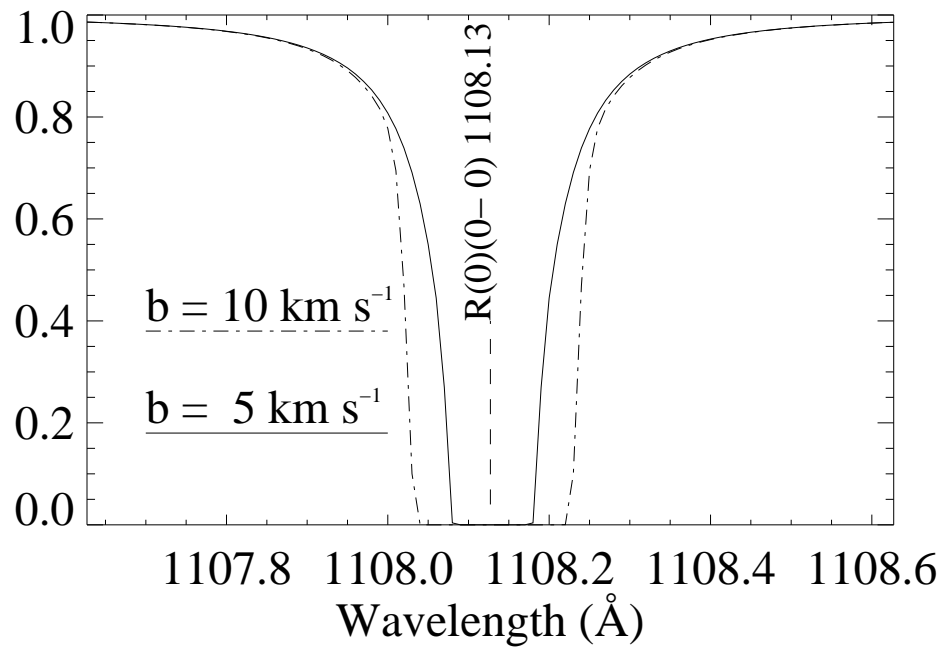
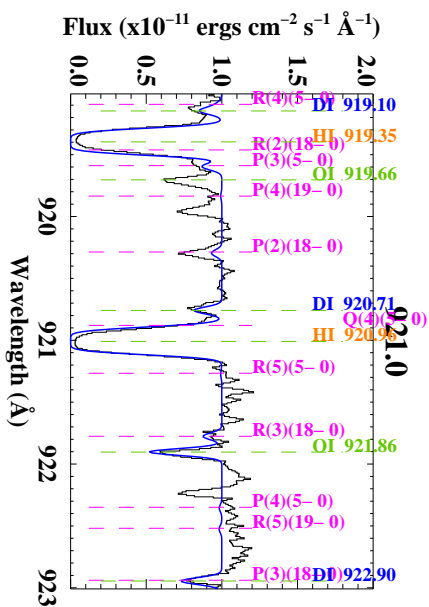
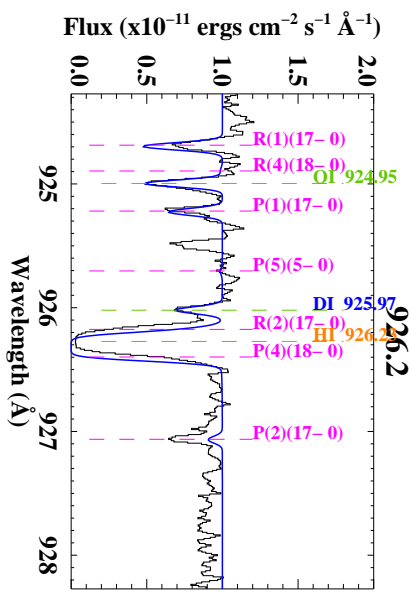
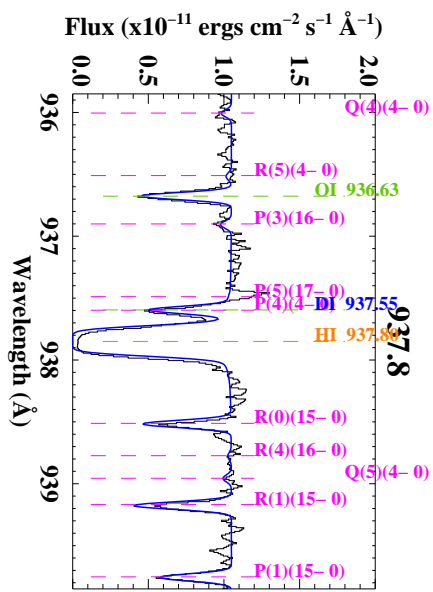
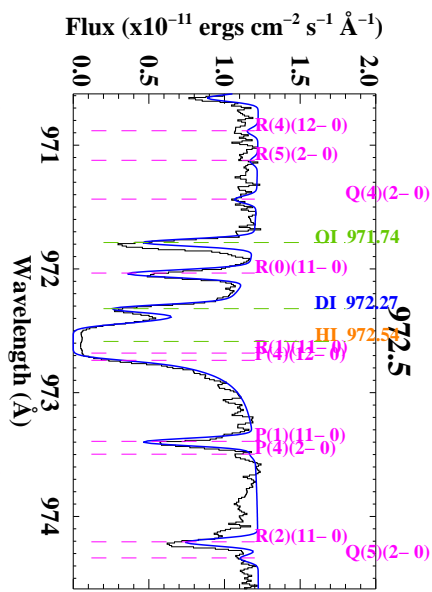
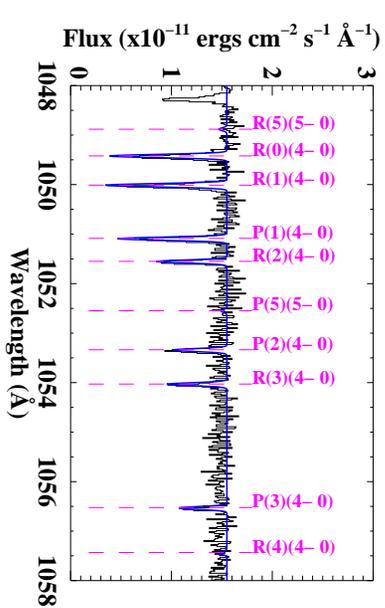
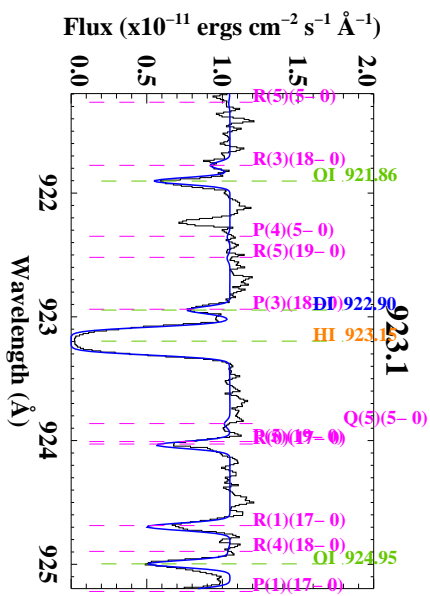
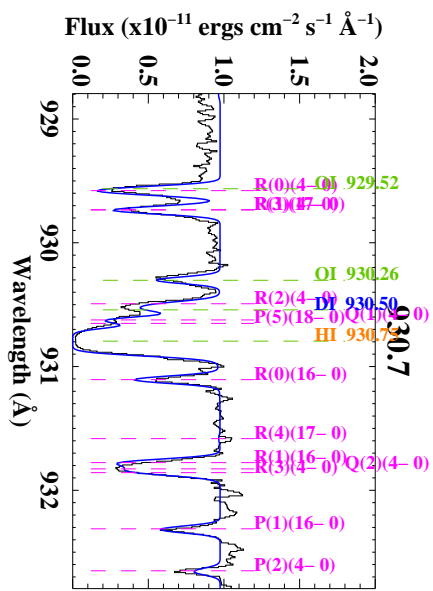
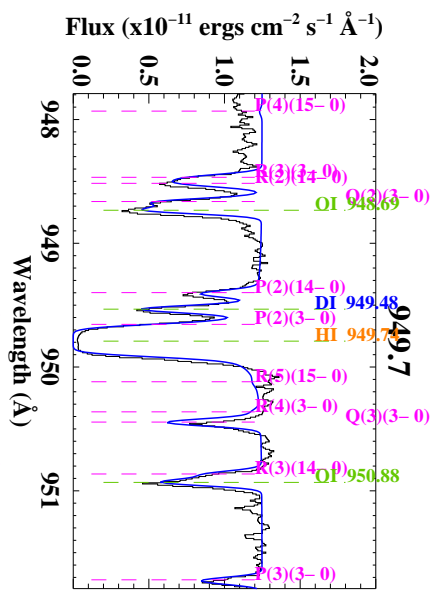
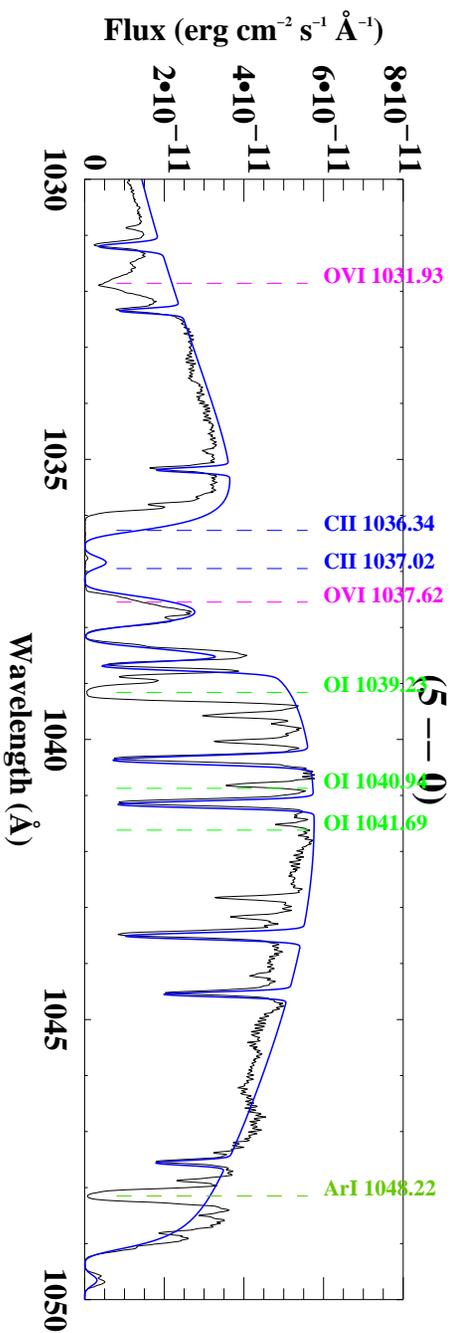
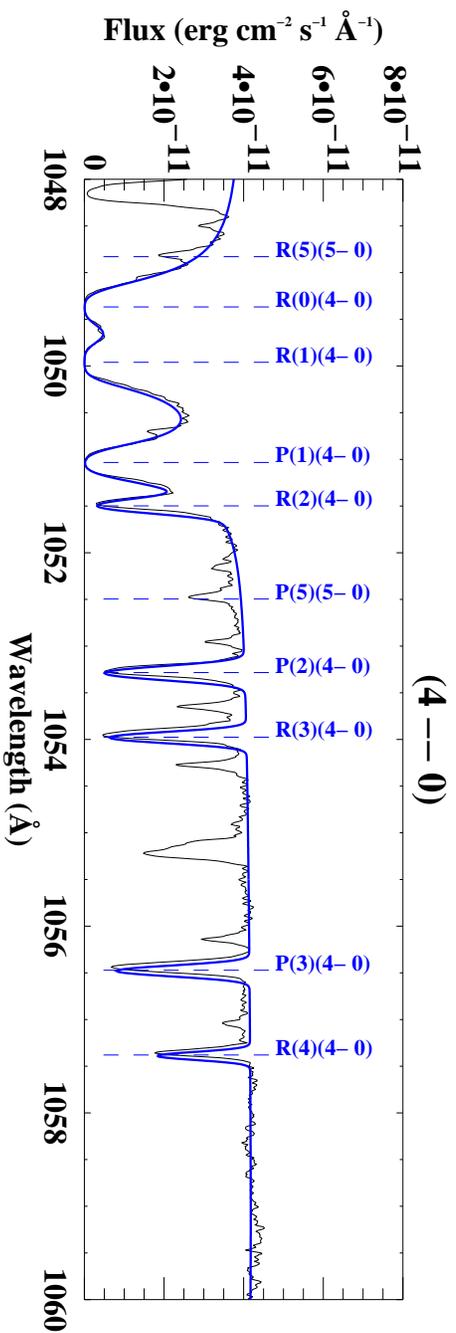
\end{document}